\documentclass{article}
\usepackage{amsmath,amsfonts}
\usepackage{graphicx}
\usepackage{natbib}
\usepackage{url}
\usepackage{algorithm}
\usepackage{algpseudocode}
\usepackage{float}
\usepackage{caption}
\usepackage{bigstrut}

\makeatletter
\algnewcommand{\LineComment}[1]{\Statex \hskip\ALG@thistlm
 $\triangleright$ #1}
\algnewcommand{\IndentLineComment}[1]{\Statex \hskip\ALG@tlm
 $\triangleright$ #1}
\makeatother

\begin{document}

\def\half{{\scriptstyle\frac{1}{2}}}

%%%%%%%%%%%%%%%%%%%%%%%%%%%%%%%%%%%%%%%%%%%%%%%%%%%%%%%%%%%%%%%%%%%%%%%%%%%%%%

\title{\textbf{Perfect simulation from unbiased simulation}}
\author{George M. Leigh$^{1}$*, Wen-Hsi Yang$^{2,3}$,\\
 Montana E. Wickens$^1$ and Amanda R. Northrop$^4$\\
 \hspace{0.2cm}\\
 $^1$Fisheries Queensland, Department of Agriculture and Fisheries\\
 41 George St., Brisbane Qld 4000, Australia\\
 $^2$School of Mathematics and Physics, University of Queensland\\
 St Lucia, 4072, Queensland, Australia\\
 $^3$School of Agriculture and Food Sciences, University of Queensland\\
 St Lucia, 4072, Queensland, Australia\\
 $^4$Fisheries Queensland, Department of Agriculture and Fisheries\\
 Mayers Road, Nambour Qld 4560, Australia\\
 \hspace{0.2cm}\\
 *Corresponding author email: george.leigh@daf.qld.gov.au,
  george.m.leigh@gmail.com\\
 Contributing authors: w.yang@uq.edu.au;\\
 montana.wickens@daf.qld.gov.au; amanda.northrop@daf.qld.gov.au\\
 \hspace{0.2cm}\\
 ORCiDs: GL 0000-0003-0513-2363, WY 0000-0001-8236-0082,\\
  MW 0000-0002-7316-014X, AN 0000-0002-8661-2459\\
}
\maketitle

% LENGTH GUIDES: ABSTRACT 200 WORDS, KEYWORDS 3 TO 6
% We have 193 words and 6 keywords.  The final result for the normal mixture
%   can be omitted if required.  We have used only keywords that don't appear
%   in the title of the paper.

\bigskip
\begin{abstract}
 We show that any application of the technique of unbiased simulation becomes
  perfect simulation when coalescence of the two coupled Markov chains can be
  practically assured in advance.  This happens when a fixed number of
  iterations is high enough that the probability of needing any more to
  achieve coalescence is negligible; we suggest a value of $10^{-20}$.  This
  finding enormously increases the range of problems for which perfect
  simulation, which exactly follows the target distribution, can be
  implemented.  We design a new algorithm to make practical use of the high
  number of iterations by producing extra perfect sample points with little
  extra computational effort, at a cost of a small, controllable amount of
  serial correlation within sample sets of about 20 points.  Different sample
  sets remain completely independent.  The algorithm includes maximal coupling
  for continuous processes, to bring together chains that are already close.
  We illustrate the methodology on a simple, two-state Markov chain and on
  standard normal distributions up to 20 dimensions.  Our technical
  formulation involves a nonzero probability, which can be made arbitrarily
  small, that a single perfect sample point may have its place taken by a
  ``string'' of many points which are assigned weights, each equal to $\pm 1$,
  that sum to~$1$.  A point with a weight of $-1$ is a ``hole'', which is an
  object that can be cancelled by an equivalent point that has the same value
  but opposite weight $+1$.
\end{abstract}

\noindent%
\vfill

\newpage

\section{Introduction}\label{sec:intro}

Monte Carlo methodology has become an indispensable numerical tool in science \citep{liu_monte_2004}, most notably in the highly versatile form of Markov chain Monte Carlo (MCMC) \citep{brooks_handbook_2011}.  MCMC uses random numbers to iterate a Markov process and produce chains of simulated random variables whose distributions gradually converge to a desired target distribution.  A critical part of MCMC practice, therefore, comprises the evaluation of MCMC convergence.

Numerous numerical and visual tools for diagnosing MCMC convergence are commonly applied \citep[see][and references therein]{roy_convergence_2020}.  None of these tools can detect convergence exactly in practice and show itself consistently superior to others; hence, practitioners often use several of them together.  In addition, MCMC generally produces serially correlated samples, leading to slow convergence and ending in poor approximation to the target distribution.  As a result, it is common to run independent chains for as many iterations as computationally feasible, and use thinning \citep{gelman_inference_2011} to reduce serial correlation, although this results in discarding most of the generated MCMC sample points.

The development of unbiased simulation \citep{glynn_exact_2014, jacob_unbiased_2020} has shown that coupling two Markov chains can produce an unbiased simulation of target distributions without excessive computation. In the simulation, the two coupled chains $X$ and~$Y$ have the same distribution of starting points, which are generated independently, after which their random numbers (which are used, for example, to generate a proposed sample point for the next MCMC step and to decide whether to accept or reject it) are offset by one iteration.  The transition from $Y_0$ to $Y_1$ uses the same random numbers as that from $X_1$ to $X_{2\,}$, etc.  Unbiased simulation is computationally effective in requiring only two chains to be simulated and not requiring the amount of iteration to be known in advance.  Simulation can be halted immediately upon coalescence when the use of the same random numbers makes $X_i = Y_{i - 1}$ for some index~$i$.

The flexibility of unbiased simulation has allowed it to be implemented in sequential Monte Carlo \citep{van_den_boom_unbiased_2022} and in Hamiltonian Monte Carlo (HMC) \citep{heng_unbiased_2019,leigh_design_2022}.  Its use in HMC is especially attractive and offers the capability, for quite general continuous processes, to completely overcome the convergence difficulties that plague current usage of MCMC.  We will return to HMC in the Discussion section of this paper.

The disadvantage of unbiased simulation is that its output consists not of coalesced sample values but of potentially lengthy sums that can vary wildly and even produce values outside the allowed sample space, such as estimated probability values that are negative or greater than~1. 

Perfect simulation overcomes the above drawback of unbiased simulation by generating samples that exactly follow the target distribution and are independent \citep[see reviews by][]{craiu_perfection_2011, huber_perfect_2016}.  Currently, few problems are amenable to perfect simulation, whereas a much greater range of problems is suited to unbiased simulation.

The most versatile existing algorithm that can achieve perfect simulation is coupling from the past (CFTP) \citep{propp_exact_1996}, but even that algorithm’s formal accomplishment of perfect simulation is limited to a small subset of the total range of problems suitable for MCMC.  CFTP uses coupling to produce coalescence from different starting points, and constitutes perfect simulation when it can be proven that the same outcome occurs from every possible starting point.  CFTP samples forward in time but achieves coalescence by prepending extra random numbers as needed to the beginning of the sequence, not the end, thereby avoiding bias towards ``easier'' sets of random numbers that yield faster coalescence.  The variant read-once coupling from the past (ROCFTP) \citep{wilson_how_2000} proceeds only in forward sequence, in blocks of fixed length, some of which exhibit coalescence while others may not.

Proof of coalescence from every possible starting point in CFTP can be achieved through a \textit{monotonicity} property which is possessed by some processes \citep{propp_exact_1996}.  In general, however, monotonicity is not available.  Even a discrete-valued process whose transition probabilities are \textit{a priori} not known numerically but are calculated on the run would have to visit every possible state to establish perfect simulation with CFTP or ROCFTP \citep[p.~135]{lovasz_exact_1995, fill_interruptible_1998}.  Continuous processes with non-countable states magnify this difficulty.

Diverse perfect simulation algorithms either build on CFTP or are constructed independently of CFTP.  Notable examples include multigamma coupling~\citep{murdoch_exact_1998}, perfect slice sampling~\citep{mira_perfect_2001} and strong stationary stopping times~\citep{aldous_strong_1987}.  These algorithms are generally difficult to use and not widely applicable.  Multigamma coupling has an impossibly small probability of satisfying the conditions for perfect simulation for most practical problems.  Perfect slice sampling has the difficulty of establishing, at least approximately, the boundary of the feasible space, which leads to a curse of dimensionality \citep{bellman_review_1955} in high-dimensional spaces.

In this paper, we will show in section~\ref{sec:unbiasedperfect} that any application of unbiased simulation becomes perfect simulation when coalescence can be practically assured in advance, i.e., when a high enough, fixed number of iterations are performed that the probability of needing any more to achieve coalescence is negligible.  This finding enormously increases the range of problems for which perfect simulation can be implemented.  In section~\ref{sec:algorithm}, we design a new algorithm to make practical use of the extra computation, to produce extra perfect sample points.  In section~\ref{sec:examplesimple}, we illustrate the methodology on two example processes, one discrete and one continuous.

\section{From unbiased to perfect simulation} \label{sec:unbiasedperfect}

\subsection{Recap of unbiased simulation} \label{sec:recap}

Unbiased simulation is formulated by two coupled Markov chains that define the following random variable $G$ \citep[see][p.~545]{jacob_unbiased_2020}:
\begin{equation}
 G = g(X_k) + \sum_{i = k + 1}^{\infty} \big\{ g(X_i) - g(Y_{i - 1}) \big\}, \label{eq:unbiasedsum}
\end{equation}
where $X$ and $Y$ are chains from the same homogeneous Markov process, whose starting points $X_0$ and $Y_0$ are independent and identically distributed; $g$ is a pre-defined, fairly arbitrary function whose expectation $\mathbb E(g(X_i))$ exists for $i \ge 0$ and as $i \to \infty$; and $k \ge 0$ is a pre-chosen number of iterations akin to burn-in.  Because $X_0$ and~$Y_0$ are identically distributed, and $X$ and~$Y$ follow the same Markov process, $X_i$ and~$Y_i$ are also identically distributed for $i > 0$.

The same random numbers are used in $X$ and~$Y$, but they are offset by one iteration so that those for $X_i$ are used for $Y_{i - 1}$.  Chains $X$ and~$Y$ are designed to coalesce, with probability~1, to the same values, again offset by one iteration: after some random number of iterations~$\tau$, $X_\tau = Y_{\tau - 1}$.  Because $X$ and~$Y$ are homogeneous Markov chains whose transitions depend only on the current state, they remain coalesced, so that $X_i = Y_{i - 1}$ for all $i \ge \tau$.  We emphasize that, despite this coalescence, it is $X_i$ and~$Y_i$ (with no offset) that are identically distributed.  The offset variables $X_i$ and~$Y_{i - 1}$ are, in general, not identically distributed for any fixed value of~$i$, because the time to coalesce, $\tau$, has no fixed upper bound.

Due to the coalescence, the infinite sum in (\ref{eq:unbiasedsum}) converges and any realization of it contains, with probability~1, only a finite, although random, number of terms.  Being finite, the sum can be calculated exactly, subject only to available numerical precision, with no need for asymptotic approximations.

Because $X$ and~$Y$, this time with no offset in the subscript, are identically distributed, $\mathbb E(g(Y_{i - 1}))$ can be replaced by $\mathbb E(g(X_{i - 1}))$ in the expectation of the random variable $G$ in (\ref{eq:unbiasedsum}), which becomes
\begin{equation}
 \begin{split}
  \mathbb E(G) &= \mathbb E(g(X_k)) + \sum_{i = k + 1}^{\infty} {\big \{\mathbb E(g(X_i)) - \mathbb E(g(X_{i - 1})) \big\}} \\
  & = \lim_{i \to \infty} \mathbb E(g(X_i)) = \mathbb E(g(Z)), \label{eq:EGEZ}
 \end{split}
\end{equation}
where the random variable $Z$ is drawn from the stationary distribution of $X$ and~$Y$.  Therefore $G$ is an unbiased simulation of the distribution of~$g(Z)$.  We do not refer to it as a ``sample'' because it may lie outside the range of~$g(Z)$.

It is essential that the coalescence after $\tau$ iterations occur for the original chains $X$ and~$Y$: it is not sufficient that $g(X_\tau) = g(Y_{\tau - 1})$, because then it may be possible for $g(X)$ and~$g(Y)$ to later diverge.

\subsection{High probability of a perfect sample} \label{sec:highprob}

It is important that this section be taken in combination with the next section~\ref{sec:relaxed}.  We state below that the probability that the theory in section~\ref{sec:relaxed} will be required in practice can be made arbitrarily small, but this does not mean that the event it covers can be neglected.  To establish perfect sampling, that theory remains critical.

We postulate a random variable $Q$ that does not depend on the function~$g$ and whose conditional expectation satisfies, for any choice of~$g$,
\begin{equation}
 \mathbb E(g(Q) \mid X, Y) = G = g(X_k) + \sum_{i = k + 1}^{\infty} \big\{ g(X_i) - g(Y_{i - 1}) \big\}. \label{eq:postulate}
\end{equation}
In this section we show that such a random variable, if it exists, is a perfect sample from the target distribution, and that, when a large value is chosen for the burn-in length $k$, there is a high probability that $Q = X_k$.  The form of $Q$ when $Q \ne X_k$ is discussed in the next section~\ref{sec:relaxed}.

Consider the case of taking $g = I_A$, an indicator function defined on an arbitrary measurable subset~$A$ of the range of $X$ and~$Y$.  Then, from (\ref{eq:EGEZ}) and~(\ref{eq:postulate}),
\begin{equation}
 \mathbb E(I_A(Q)) = \mathbb E(G) = \mathbb E(I_A(Z)) = \mathbb P(Z \in A). \label{eq:expecindicator}
\end{equation}
This equation holds for any choice of~$A$; and $Q$, as postulated, does not depend on $g$ or~$A$.  The probability distributions of $Q$ and~$Z$ are completely determined by the probabilities that these variables belong to an arbitrary subset of their range.  Therefore, (\ref{eq:expecindicator}) implies that these distributions must be the same; i.e., if a random variable $Q$ can be found that satisfies~(\ref{eq:postulate}) for all choices of~$g$, it must be a perfect sample from the stationary distribution of $X$ and~$Y$.

If the number of iterations $\tau$, at which $X$ and~$Y$ coalesce, is less than or equal to $k + 1$, $Q$ is simply equal to $X_k\,$, as the sum in~(\ref{eq:postulate}) is then empty.

Therefore, setting $k$ to a high quantile of~$\tau$ results in a high probability that $X_k$ is a perfect sample, provided that $Q$ can still be defined when $\tau > k$.  We explore the latter case in the next section.  The probability $\mathbb P(\tau \le k)$ can be made arbitrarily close to~1 by choosing a high enough value for~$k$.

\subsection{Relaxed perfect sampling} \label{sec:relaxed}

However large $k$ is chosen, the probability that $\tau > k + 1$, i.e., that $X$ and~$Y$ do not coalesce within $k + 1$ iterations, will still be nonzero.  Here we describe what to do when this occurs.

We set $Q$ to an improper conditional distribution which we call a ``string'' of values~$q_i$ and associated weights~$w_i$:
\begin{equation}
 \begin{split}
  Q &= ((q_{1\,}, w_1), (q_{2\,}, w_2),\, \ldots\,) \\
  &= ((X_{k\,}, 1), (Y_{k\,}, -1), (X_{k + 1\,}, 1), (Y_{k + 1\,}, -1),\,
   \ldots\,) \label{eq:stringdef}
 \end{split}
\end{equation}
where a weight of~$1$ signifies a conventional sample point, while a weight of $-1$ signifies an improper sample point which we call a ``hole''.  A hole is defined as an object that would be cancelled by a conventional sample point with the same value $q$ but opposite weight $w = 1$.  It is similar to a positively charged hole in chemistry which does not exist as an actual particle but is cancelled by a negatively charged electron that moves into the vicinity.

As in~(\ref{eq:unbiasedsum}), due to the coupling of $X$ and~$Y$, the sequence in a realization of~(\ref{eq:stringdef}) is finite, terminating after $\nu = 2(\tau - k) - 1$ terms at the final element ($X_{\tau - 1}$, 1).  Therefore a realization of~$Q$ can be calculated exactly, in the same way as a realization of~$G$ in~(\ref{eq:unbiasedsum}).  The weights within a string satisfy $w_i = (-1)^{i - 1}$ and $\sum_{i = 1}^\nu {w_i} = 1$.

We define the conditional expectation of a function of~$Q$ in the same way as for a discrete distribution, with the weights $w_i$ in place of a probability function:
\begin{equation}
 \mathbb E(g(Q) \mid X, Y) = w_1\, g(q_1) + \sum_{i = 1}^\infty {\big\{ w_{2 i}\, g(q_{2 i}) + w_{2 i + 1}\, g(q_{2 i + 1}) \big\}}.
\end{equation}
Therefore, $Q$ satisfies~(\ref{eq:postulate}), as the $w$'s are equal to $\pm 1$ and the $q$'s are equal to $X$ and~$Y$.  By the theory in section~\ref{sec:highprob}, $Q$ is, in a relaxed sense due to the presence of negative weights, a perfect sample.

The presence of holes in a sample does not affect the validity of~(\ref{eq:expecindicator}), but does affect the dispersion of the sample mean of $I_A(Q)$ (see examples in the final column of Table~\ref{tab:results2state} in section~\ref{sec:exampletwostate} below).  When $Q$ is a string consisting of both conventional sample points and holes, it contributes $\nu$ points instead of only one.  Therefore, its presence or absence in the sample will make a bigger difference to the sample mean than a single point would make.  If strings with $\nu > 1$ are rare due to a high choice for~$k$, their effect on the overall precision of estimation will be small.

\subsection{Comments}

We stress that the target distribution should be explored both theoretically and numerically, using a widely dispersed distribution of starting points, before the value of $k$ is set.  Consequences of the presence of holes can be severe: a single string can be very long and contain many holes, e.g., when the target distribution is multi-modal, with one region of the sample space almost cut off from the rest, and this is not appreciated before setting~$k$.

In the following section, we present methodology to reduce the computational effort per perfect sample point, at the expense of a small amount of serial correlation within relatively small sample sets of typically 20 sample points.  Different sample sets remain independent.

\section{Practical algorithm for perfect simulation} \label{sec:algorithm}

\subsection{Algorithm structure} \label{sec:algstructure}

The theory in section~\ref{sec:unbiasedperfect} achieves perfect simulation by setting the number of iterations, $k$ in equation~(\ref{eq:unbiasedsum}), very high.  In this section, we present a practical algorithm to reduce the amount of computation.

The algorithm groups sample points into sample sets of size~$K$, within which sample points are allowed to have small correlations. A thinning ratio or block length, $B$, is chosen which determines the magnitude of the internal correlations. The burn-in number of iterations is set to $k = K B$.  We offset the coupled chains $X$ and~$Y$ in~(\ref{eq:unbiasedsum}) by $B$ iterations instead of just one iteration, and run each chain for $K$ blocks each of $B$ iterations, noting that the process whose ``iteration'' is a block of $K$ iterations of $X$ or~$Y$ is still a Markov process.  Equation~(\ref{eq:unbiasedsum}) is modified to
\begin{equation}
 G = g(X_{KB}) + \sum_{i = K + 1}^{\infty} \big\{ g(X_{iB}) - g(Y_{(i - 1)B}) \big\}. \label{eq:unbiasedsummod}
\end{equation}

Parameter $B$ is chosen to regulate the probability that the chains coalesce within one block:
\begin{equation}
 \mathbb P(X_{2B} = Y_B) \approx 1 - P, \label{eq:probcoalblock}
\end{equation}
where the starting points $X_0$ and~$Y_0$ are chosen independently from a widely dispersed distribution that is thought to cover the range of $X$ and~$Y$, and $P$ is an acceptable probability of lack of coalescence after $B$ iterations.  A nonzero probability of non-coalescence allows some correlation between chains in the same sample set.  This can be reduced by using a smaller value of $P$, at the expense of fewer points in the sample set for the same amount of computation.  Equation~(\ref{eq:probcoalblock}) is an approximation because the value of $B$ as a function of~$P$ generally cannot be found analytically and has to be determined by a preliminary simulation experiment.

The sample-set size $K$ is determined by the desired probability of non-coalescence after $K + 1$ blocks of $B$ iterations.  With widely dispersed starting points, continuing from the end of the previous block should be no less successful than running from the original starting points.  Hence the probability of non-coalescence can be expected to decay geometrically, reaching approximately $P^K$ after $K + 1$ blocks.  Lack of coalescence after $K + 1$ blocks produces holes in the results, as described in section~\ref{sec:relaxed}.  A larger choice of $K$ reduces the probability of occurrence of holes.

We suggest that $P$ can be set to 0.1 and $K$ around 20, to produce a probability of a hole of about~$10^{-20}$, provided that the sample space has been thoroughly analysed and the distribution of starting points is widely dispersed.  Higher values of~$K$ can be used if desired.  The value of $K$ should also be increased according to the total number of sample points to be generated, $n$, so that the probability of getting any holes within the entire set of results, approximately $n P^K$, is kept very low.

\begin{figure}[ht]
 \centering
 \includegraphics[width=5.25cm, trim=2.1cm 7.5cm 7cm 1.5cm,
  clip=true]{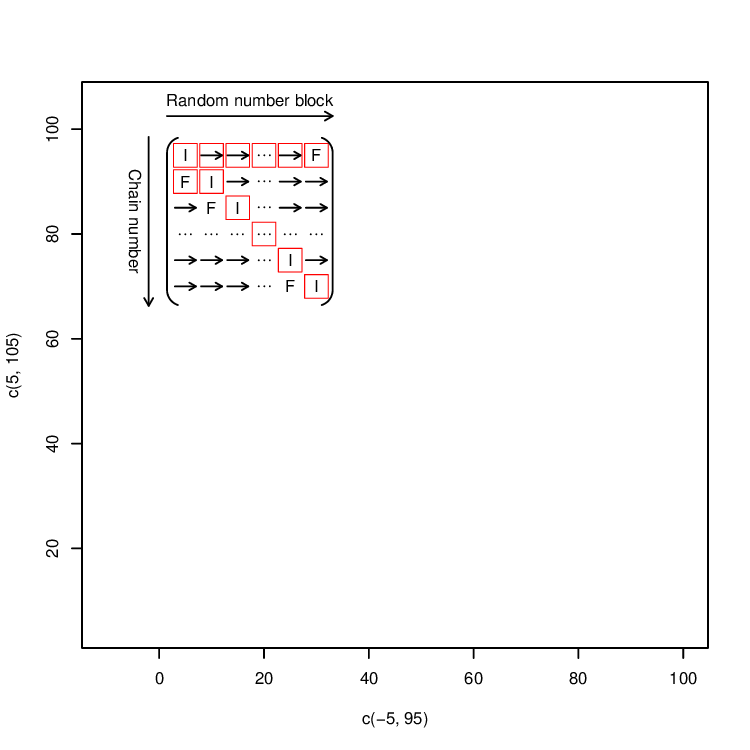}
 \caption[Chain $\times$ block matrix design]
 {Chain $\times$ block matrix for one sample set.  Random-number blocks proceed from left to right and are common to all chains.  Each chain is initialized on its diagonal element ``I'' to a starting point from some preset distribution.  Blocks and chains wrap around when necessary.  The final cell ``F'' in each chain is a perfect sample if it coalesces with the chain below it.  Cells marked by red squares are always executed; others only if chains haven't yet coalesced.}
  \label{fig:sampleset}
\end{figure}

Our algorithm is shown in Fig.~\ref{fig:sampleset}, in the form of a $K \times K$ matrix, and in Algorithm~\ref{alg:practical}.  Each element of the matrix corresponds to a block of $B$ MCMC iterations.  Initialization to a starting point takes place at the beginning of each block on the diagonal of the matrix.  Blocks are then applied from left to right, the input to each off-diagonal block being the output from the block to the left of it.  Blocks wrap around back to column~1 after reaching column~$K$, and, with high probability, produce a perfect sample at the end of each block immediately to the left of the diagonal.  For purposes of illustration, Algorithm~\ref{alg:practical} shows the result variable $Q$, which represents a column of Fig.~\ref{fig:sampleset}, as a vector.  For a $d$-dimensional process with $d > 1$, it is actually coded as a $K \times d$ matrix.  The same is true for the variable $Q_1$ which stores the first row of Fig.~\ref{fig:sampleset}.

% GML: I have replaced "\Leftarrow" from the Springer-Nature template by "\gets".  I can't stomach the double arrow.  To me, it means "is implied by", not "is assigned".  Max line number in algorithm = 35.  Otherwise algorithm goes to end of paper.

\begin{algorithm}[!th]
 \caption{Practical algorithm for one perfect sample set of size~$K$} \label{alg:practical}
 \begin{algorithmic}[1]
  \Require $K \ge 1$, $B \ge 1$, functions \Call{Start}{$K$}, \Call{Rand}{$B$}, \Call{Mcmc}{$X_0$, $R$, $B$}.
  \Ensure $Q$ stores $K$ perfect sample points from \Call{Mcmc}{}, if Error = \textbf{false}.
  \Function{MinInd}{$x, i_1, i_2$} \Comment{Index of closest of $x[i_1:(i_2 - 1)]$ to $x[i_2]$}
   \State \Return $\Call {WhichMin} {\big\lvert x[i_2] - x[i_1:\{i_2 - 1\}] \big\rvert} + i_1 - 1$ \Comment{Metric required} \label{line:metric}
  \EndFunction \Comment{\Call {WhichMin}{}: index of minimum, first one if there are ties} \label{line:endminind}
  \State $Q \gets$ \Call{Start}{$K$} \Comment{Vector of random starting points on diagonal of Fig.~\ref{fig:sampleset}}
  \State $Q_1 \gets \textbf{zeros}(K)$ \Comment{Row vector to store first row of Fig.~\ref{fig:sampleset} for later use.}
  \State $s \gets$ \Call{Seed}{}; Error $\gets$ \textbf{false} \Comment{Store random seed; set coalescence error.}
  \State $a \gets \textbf{zeros}(K)$ \Comment{Mark all chains as active (not coalesced yet).}
  \For{$j \gets 1:K$} \Comment{Loop over blocks (columns) in Fig.~\ref{fig:sampleset} upper triangle.} \label{line:uppertri1}
   \State $R \gets$ \Call{Rand}{$B$} \Comment{Generate random numbers for $B$ MCMC iterations.} \label{line:randorig}
   \For{$i \gets 1:j$} \Comment{Loop over chains (rows) in Fig.~\Ref{fig:sampleset} upper triangle.}
    \If {$a[i] > 0$} $Q[i] \gets Q[a[i]]$ \textbf{else} \Comment{Copy if already coalesced.}
     \State $Q[i] \gets$ \Call{Mcmc}{$Q[i], B, R$} \Comment{MCMC from current value of $Q$} \label{line:upperreplacefirst}
     \If {$i = 1$} $Q_1[j] \gets Q[1]$ \textbf{else} \Comment{Store first row of Fig.~\ref{fig:sampleset}.}
      \State $m \gets \Call {MinInd} {Q, 1, i}$ \Comment{Closest previous value in column~$j$} \label{line:setmupper}
      \State \textbf{if} {$Q[i] = Q[m]$} \textbf{then} $a[i] \gets m$ \textbf{end if} \Comment{Record coalescence.}
     \EndIf \Comment{Nonzero $a[i]$ records row with which row~$i$ coalesced.} \label{line:upperreplacelast}
    \EndIf
   \EndFor \Comment{$i$ loop (rows of Fig.~\ref{fig:sampleset})} \label{line:endiloop}
  \EndFor \Comment{$j$ loop (columns of Fig.~\ref{fig:sampleset})} \label{line:uppertri2}

  \State \Call{Seed}{} $\gets s$ \Comment{Restore seed to regenerate previous random number blocks.} \label{line:lowertri1}

  \For{$j \gets 1:(K - 1)$} \Comment{Loop over blocks in Fig.~\ref{fig:sampleset} lower triangle.}
   \State Error $\gets$ Error \textbf{or} $a[j + 1] \ne j$ \Comment{Check final coalescence of chain~$j$.}
   \State $a[j + 1] \gets a[j]$; $a[a = j] \gets j + 1$ \Comment{Move coalescence from row~$j$ to~$j + 1$.}
   \State $Q[1] \gets Q_1[j]$ \Comment {Reload row 1 saved copy; no need to recompute.} \label{line:reloadrow1}
   \State $R \gets$ \Call{Rand}{$B$} \Comment{Regenerate random numbers for block~$j$.}
   \For{$i \gets (j + 1):K$} \Comment{Loop over chains in Fig.~\Ref{fig:sampleset} lower triangle.}
    \If {$a[i] > 0$} $Q[i] \gets Q[a[i]]$ \textbf{else} \Comment{Copy, else run MCMC.}
     \State $Q[i] \gets$ \Call{Mcmc}{$Q[i], B, R$} \Comment{Run MCMC \& check coalescence.} \label{line:mcmclower}
     \State \textbf{if} $i = j + 1$ \textbf{then} $m \gets 1$ \textbf{else} $m \gets \Call {MinInd} {Q, j + 1, i}$ \textbf{end if} \label{line:setmlower}
     \State \textbf{if} $Q[i] = Q[m]$ \textbf{then} $a[i] \gets m$ \textbf{end if} \Comment{Coalesced with row~$m$}
    \EndIf \Comment{Lower triangle of Fig.~\ref{fig:sampleset} concerns rows 1 and $(j + 1):i$}
   \EndFor \Comment{$i$ loop (rows of Fig.~\ref{fig:sampleset})}
  \EndFor \Comment{$j$ loop (columns of Fig.~\ref{fig:sampleset})}
  \State Error $\gets$ Error \textbf{or} $a[K] \ne 1$; $Q[1] \gets Q_1[K]$ \Comment{Check $Q[K]$, restore $Q[1]$.} \label{line:lowertri2}

 \end{algorithmic}
\end{algorithm}

Each column of the matrix in Fig.~\ref{fig:sampleset} represents a block of random numbers: the same random numbers are used in that column in every row, to couple each row with the one below it.  Then each row provides a perfect sample point after $K$ blocks if it has coalesced with the row below at that time, or if it coalesces after one extra block with new random numbers, making $K + 1$ blocks for the upper row and $K$ blocks for the lower row; the two rows are used as the processes $X$ and~$Y$ in (\ref{eq:unbiasedsummod}).  Chain~$K$ is required to coalesce with row~1 in column $K - 1$, or after one extra block with new random numbers.

Algorithm~\ref{alg:practical}, line~\ref{line:metric}, with continuous processes in mind, requires a metric to measure the relative closeness of one chain to another, and hence (lines \ref{line:setmupper} and~\ref{line:setmlower}) to choose the previous chain most likely to coalesce with the current chain.  For continuous processes as discussed in section~\ref{sec:maximal}, this is intended to be the Euclidean metric, i.e., the usual absolute value function.  For discrete processes with unordered states, a simple zero-one function suffices, taking the value zero when then states are equal and 1 when they are different.

Extra iteration to that detailed in Algorithm~\ref{alg:practical}, using new random numbers as needed, is required to calculate the right-hand side of~(\ref{eq:unbiasedsummod}) if a row does not coalesce with the row below it within $K$ blocks.  Holes occur if the two rows still do not coalesce after $K + 1$ blocks.
 
We emphasize that different sample sets remain completely independent.  Correlations occur only within the same sample set of $K$ points.

Our algorithm produces a sample set of $K$ points with similar computational effort to that for two independent perfect sample points.  Most of the chains coalesce after a single block, as determined by the probability of non-coalescence $P$ in~(\ref{eq:probcoalblock}).  After a row has coalesced with a row above it in Fig.~\ref{fig:sampleset}, computation for the lower row can be curtailed and its results set to those for the higher row.  Computation always has to be conducted for the cells marked by red squares: along row~1, the diagonal, and block~1 in row~2.

The algorithm can be parallelized easily by assigning different sample sets, which are unconnected and contain roughly 20 points, to different processors.

\subsection{Maximal coupling for continuous processes} \label{sec:maximal}

Algorithm~\ref{alg:practical} is designed to facilitate maximal coupling in simulation of continuous processes.  Maximal coupling originated as a method of analysis of convergence of a Markov chain to its stationary distribution, in which one of the coupled chains began from some prescribed initial distribution and the other from the stationary distribution \citep{griffeath_maximal_1975}.  It evolved into a scheme to facilitate coalescence of any two chains whose probability densities overlap \citep[see, e.g.,][]{johnson_coupling-regeneration_1998}.  

In our setting, maximal coupling inserts into each chain a proposed jump to a new destination point, which is accepted with some probability according to a Metropolis--Hastings (M--H) update.  The destination points from two different chains can, with a certain probability given by the overlap of their probability densities, be made equal while still following their preset jump distribution.  An M--H update \citep{metropolis_equations_1953,hastings_monte_1970} accepts a move from an existing point $X$ to a proposed new point $X^*$ with probability $\min(\exp\{U(X) - U(X^*)\}, 1)$, where $U$ is the negative log-likelihood function.

It is sometimes feasible to maximally couple chains directly in their internal MCMC, without the need for a separate M--H update.  The method in this section is general, making no assumptions about internal MCMC structure.

\begin{figure}[ht]
 \centering\includegraphics[width=8cm, trim=3cm 4cm 2.5cm 3.5cm, clip=true]{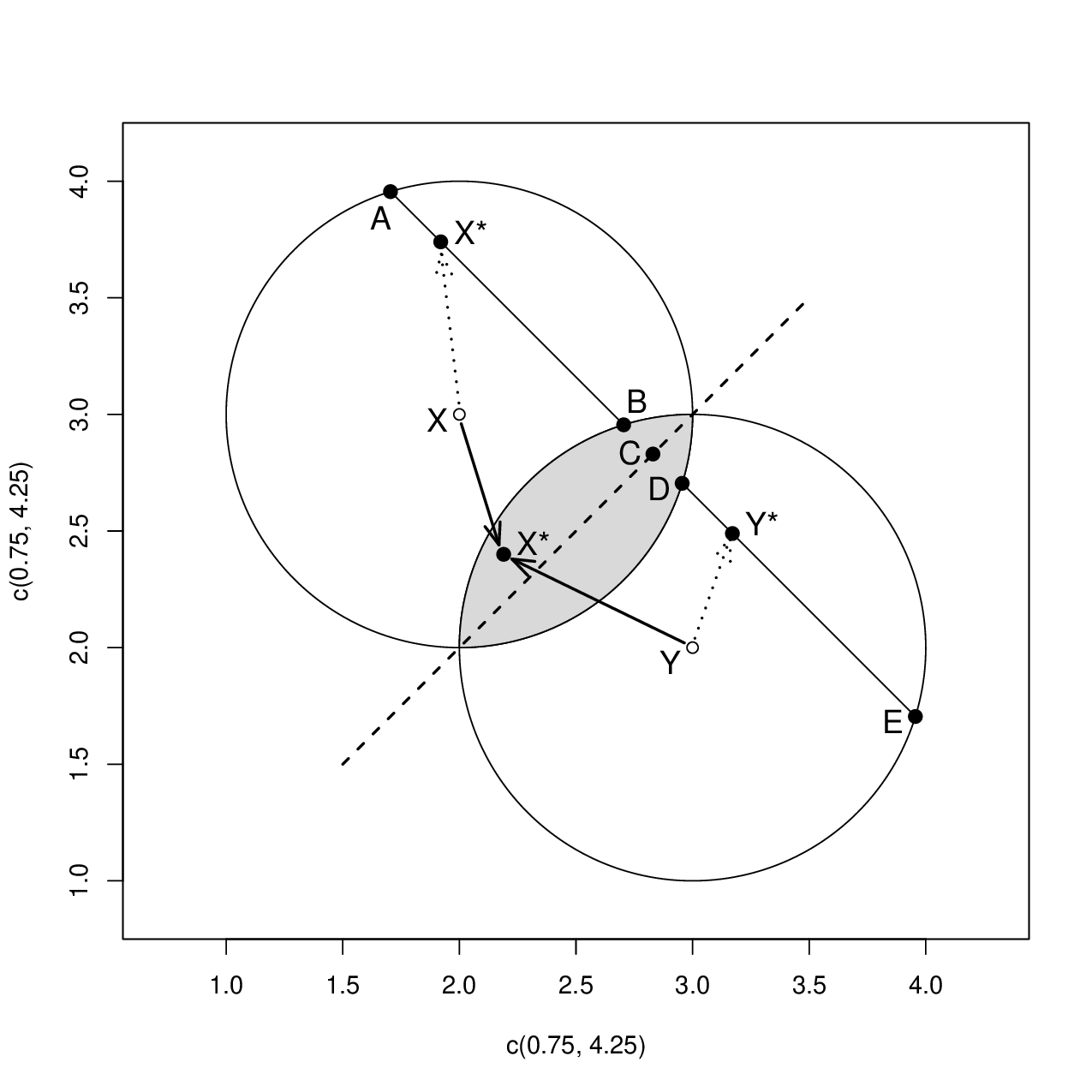}
 \caption[Maximal coupling using a solid spherical uniform distribution]
 {Maximal coupling of destination points, using a solid spherical uniform distribution for the Metropolis--Hastings jumps.  Open and solid circles show origin points $X$ and~$Y$, and proposed destination points $X^*$ and~$Y^*$, of the jumps of the two processes.  If the independently generated $X^*$ happens to lie in the overlap (shaded) zone with the jump-hypersphere of $Y$, then $Y^*$ is set to $X^*$ (solid arrows).  Otherwise, line AE passes through $X^*$ parallel to XY, and segment AB is translated to segment DE.  Then $Y^* - D = X^* - A$, which ensures that $Y^*$ is also outside the overlap zone (dotted arrows)}
  \label{fig:maximal}
\end{figure}

We advocate a solid spherical uniform distribution with a fixed radius $r$ for maximal-coupling jumps.  In $d$ dimensions, the destination point $X^*$ is chosen uniformly over the interior of a hypersphere of radius~$r$, centred at the existing point~$X$.  We presume that, when $d > 1$, the $d$ coordinates of the process have been scaled to be non-dimensional and comparable.  The probability density function (p.d.f.) of the jump is constant throughout the hypersphere, not restricted to the surface.  Therefore the p.d.f.s of the destination points of chains $X$ and~$Y$ are identical wherever their hyperspheres overlap; see Fig.~\ref{fig:maximal}.   Overlap occurs when their centres satisfy $\lvert Y - X \rvert < 2 r$.

\begin{algorithm}[!th]
 \caption{Maximally coupled jumps for Metropolis--Hastings updates} \label{alg:maximal}
  \begin{algorithmic}[1]
   \Require $U$: negative log-likelihood function; $d$: number of dimensions.
   \Ensure $X^*, Y^*$ are maximally coupled jump destinations of $X$ and~$Y$.
   \Function{Jump}{$X, r, R_\text{dir}, R_\text{mag}$} \Comment{$X$: jump origin; $r$: radius of sphere}
   \State \Comment{$R_\text{dir}, R_\text{mag}$: random numbers: $d \times N(0, 1)$, $1 \times \text{uniform}(0, 1)$}
    \State $\hat u \gets R_\text{dir} \big/ \big\lvert R_\text{dir} \big\rvert$ \Comment{Jump direction, unit vector from Euclidean norm}
    \State $X^* \gets X + r \, R_\text{mag}^{1/d} \, \hat u$ \Comment{Spherical uniform distribution for jump}
   \State \Return $X^*$ \Comment{Destination point for~$X$, to pass to M--H acceptance test}
  \EndFunction
  \Function{MaxCouple}{$X, X^*, Y, r$} \Comment{$Y$: second jump origin}
   \State $u_X \gets X^* - X$ \Comment{Jump vector used for~$X$}
   \State $L \gets \half \lvert Y - X \rvert$ \Comment{Distance to midpoint of XY}
   \State $v \gets \half (Y - X) / L$ \Comment{Unit vector in direction of XY}
   \State $c \gets L v + u_X - (u_X \cdot v) v$ \Comment{$X$ to midpoint, plus projection of $u_X$}
   \If {$\big \lvert Y - X^* \big \rvert \le r$} \Comment{$X^*$ is in overlap zone of $X$ and $Y$ spheres.}
    \State $Y^* \gets X^*$ \Comment{Couple the two destination points.}
   \ElsIf {$\lvert c \rvert < r$} \Comment{Point~C is in overlap zone.}
    \State $w \gets -L + \sqrt{L^2 + r^2 - c \cdot c}$ \Comment{Distance from point~C to D}
    \State $u_Y \gets u_X + 2 w v$ \Comment{Translate to ensure $Y^* \notin$ overlap zone.}
    \State $Y^* \gets Y + u_Y$
   \Else \Comment{$C \notin$ overlap zone: no need for translation.}
    \State $Y^* \gets Y + u_X$ \Comment{Use same jump vector for $Y$ as for~$X$.}
   \EndIf
   \State \Return $Y^*$ \Comment{Destination point for~$Y$, to pass to M--H acceptance test}
  \EndFunction
  \Function{MhTest}{$X, X^*, R_\text{MH}$} \Comment{Metropolis--Hastings acceptance test;}
  \Statex \hskip 1.5em $R_\text{MH}$: uniform$(0, 1)$ random number
   \If {$R_\text{MH} \le \exp(U(X) - U(X^*))$}
    \State \Return $X^*$ \Comment{Accept the jump.}
   \Else
    \State \Return $X$ \Comment{Reject the jump and remain at its origin.}
   \EndIf
  \EndFunction
 \end{algorithmic}
\end{algorithm}

If the destination point $X^*$ is in the overlap zone, maximal coupling takes the same point to be $Y^*$, the destination point for $Y$.  Then $X$ and~$Y$ coalesce if both of their M--H updates are accepted.  If $X^*$ is outside the overlap zone, we must ensure that $Y^*$ is also forced outside, but still within its own hypersphere, in order to maintain the correct jump distribution for~$Y$.  A straightforward way to achieve this outcome with the solid spherical uniform jump distribution is shown by the dotted arrows in Fig.~\ref{fig:maximal}:
\begin{equation}
 u_Y = u_X + 2 w v \,, \label{eq:jumpmaximal}
\end{equation}
where $u_X = X^* - X$ and ~$u_Y = Y^* - Y$ are the jump vectors, $w = \lvert D - C \rvert $ in Fig.~\ref{fig:maximal}, and $v$ is the unit vector in the direction XY: $v = \half (Y - X) / L$ where $L = \half \lvert Y - X \rvert$.  Point C is the midpoint of XY plus the projection of $u_X$ perpendicular to XY in the dashed line in Fig.~\ref{fig:maximal}: $C = X + c$ where
\begin{equation}
 c = L v + u_X - (u_X \cdot v) v.
\end{equation}
Point D, on the surface of the hypersphere of $X$, then has position vector $D = C + w v$, so $w$ must satisfy $w > 0$ and $\lvert c + w v \rvert = r$.  Using the equality $c \cdot v = L$, the equation for~$w$ is
\begin{equation}
 c^2 + 2 L w + w^2 = r^2.
\end{equation}
For $\lvert c \rvert < r$ and $w > 0$, this equation is solved by the quadratic formula:
\begin{equation}
 w = -L + \sqrt{L^2 + r^2 - c^2} \,. \label{eq:w}
\end{equation}
If $\lvert c \rvert \ge r$, the line AE does not intersect the overlap zone, and we set $u_Y = u_X$ with no adjustment.  Our algorithm for maximal coupling is shown in Algorithm~\ref{alg:maximal}.

Updates to Algorithm~\ref{alg:practical} to enable maximal coupling are shown in Algorithms \ref{alg:replaceupper} and~\ref{alg:replacelower}.  Jumps for chain~1 are simulated freely without coupling.  In the upper triangle of Fig.~\ref{fig:sampleset} ($1 < i \le j$), during block~$j$, chain~$i$ is maximally coupled with the one of chains 1, \ldots, $i - 1$ whose pre-jump value is closest to that of chain~$i$.  In the lower triangle ($i > j$), if $i > j + 1$, chain~$i$ is maximally coupled with the one of chains $j + 1$, \ldots, $i - 1$ whose pre-jump value is closest to that of chain~$i$.  Chain~$j + 1$ is maximally coupled with chain~1 which was computed and stored when processing the upper triangle.

\begin{algorithm}[!th]
 \caption{Alg.~\ref{alg:practical} upper triangle (lines 1--\ref{line:uppertri2}) with maximal coupling} \label{alg:replaceupper}
 \begin{algorithmic}[1]
  \Require $K \ge 1$, $B \ge 1$, functions \Call{Start}{$K$}, \Call{Rand}{$B$}, \Call{Mcmc}{$X_0$, $R$, $B$}.
   \Statex Algorithm~\ref{alg:maximal}; $r$ set to maximum jump distance.
   \Statex $M =$ no. of MCMC iterations per maximal coupling step (divisor of $B$).
   \Statex $Q^-, Q_1^-$ to hold pre-jump versions of $Q$ (current column) and $Q_1$ (row~1).
   \Statex $Q^*, Q_1^*$ to hold jumped versions of $Q$ and $Q_1$, prior to M--H tests.
   \Statex $Q_1^-, Q_1^*, Q_1$ have dimensions $[K, B / M]$ to hold all max. coupling results.
   \Statex \Call{Rand}{} now returns four components: $R_\text{MCMC}$, $R_\text{dir}$, $R_\text{mag}$ and $R_\text{MH}$.
  \Ensure $Q$ returns $K$ perfect sample points from \Call{Mcmc}{}, if Error = \textbf{false}.
  \Function{MinInd}{$x, i_1, i_2$} \Comment{Index of closest of $x[i_1:(i_2 - 1)]$ to $x[i_2]$}
   \State \Return $\Call {WhichMin} {\big\lvert x[i_2] - x[i_1:\{i_2 - 1\}] \big\rvert} + i_1 - 1$ \Comment{Metric required}
  \EndFunction \Comment{\Call {WhichMin}{}: index of minimum, first one if there are ties}
  \State $J \gets B / M$ \Comment{Number of maximal coupling steps per random-number block}
  \State $Q \gets$ \Call{Start}{$K$}; $s \gets$ \Call{Seed}{}; Error $\gets$ \textbf{false}; $a \gets \textbf{zeros}(K)$
  \For{$j \gets 1:K$} \Comment{Loop over blocks (columns) in Fig.~\ref{fig:sampleset} upper triangle.}
   \For{$j_1 \gets 1:J$} \Comment{Loop over sub-blocks, each maximally coupled.}
   \State $(R_\text{MCMC}, R_\text{dir}, R_\text{mag}, R_\text{MH}) \gets$ \Call{Rand}{$M$} \Comment{Random numbers $\times M$}
   \For{$i \gets 1:j$} \Comment{Loop over chains (rows of Fig.~\ref{fig:sampleset}).}
    \If {$a[i] > 0$} $Q[i] \gets Q[a[i]]$ \textbf{else} \Comment{Copy if already coalesced.}
     \State $Q^-[i] \gets$ \Call{Mcmc}{$Q[i], M, R_\text{MCMC}$} \Comment{Run MCMC, pre-jump.}
     \If {$i = 1$} \Comment{Row~1 of Fig.~\ref{fig:sampleset}: jump freely.} \label{line:maxcoupleupper1}
      \State $Q^*[1] \gets \Call{Jump}{Q^-[1], r, R_\text{dir}, R_\text{mag}}$ \label{line:freejump}
      \State $Q_1 \gets \Call{MhTest}{Q^-[1], Q^*[1], R_\text{MH}}$
      \State $Q^-_1[j, j_1] \gets Q^-[1]$; $Q^*_1[j, j_1] \gets Q^*[1]$; $Q_1[j, j_1] \gets Q[1]$
     \Else \Comment{$i > 1$: maximally couple with closest previous row.}
      \State $m \gets \Call {MinInd} {Q^-, 1, i}$
      \State $Q^*[i] \gets \Call{MaxCouple}{Q^-[m], Q^*[m], Q^-[i], r}$ \label{line:maxcoupleupper0}
      \State $Q[i] \gets \Call{MhTest}{Q^-[i], Q^*[i], R_\text{MH}}$
      \If {$Q[i] = Q[m]$} \Comment{If coalesced, reset $m$ to earliest.}
       \State \textbf{if} $a[m] > 0$ \textbf{then} $m \gets a[m]$ \textbf{end if}
       \State $a[i] \gets m$; $a[a = i] \gets m$ \Comment{Mark row $i$, remark others.}
      \EndIf
     \EndIf \Comment{$a[i] > 0$ records row with which row~$i$ coalesced.} \label{line:maxcoupleupper2}
    \EndIf
   \EndFor \Comment{$i$ loop (rows of Fig.~\ref{fig:sampleset})}
   \EndFor \Comment{$j_1$ loop over sub-blocks}
  \EndFor \Comment{$j$ loop (columns of Fig.~\ref{fig:sampleset})}
 \end{algorithmic}
\end{algorithm}

\begin{algorithm}[!th]
 \caption{Alg.~\ref{alg:practical} lower triangle (lines \ref{line:lowertri1}--\ref{line:lowertri2}) with maximal coupling} \label{alg:replacelower}
 \begin{algorithmic}[1]
  \State \Call{Seed}{} $\gets s$ \Comment{Restore seed to regenerate previous random number blocks.}

  \For{$j \gets 1:(K - 1)$} \Comment{Loop over blocks in Fig.~\ref{fig:sampleset} lower triangle.}
   \State Error $\gets$ Error \textbf{or} $a[j + 1] \ne j$ \Comment{Check final coalescence of chain~$j$.}
   \State $a[j + 1] \gets a[j]$; $a[a = j] \gets j + 1$ \Comment{Move coalescence from row~$j$ to~$j + 1$.}
   \For{$j_1 \gets 1:J$} \Comment{Loop over sub-blocks.}
    \State $Q^-[1] \gets Q_1^-[j, j_1]$; $Q^*[1] \gets Q_1^*[j, j_1]$; $Q[1] \gets Q_1[j, j_1]$
    \State $(R_\text{MCMC}, R_\text{dir}, R_\text{mag}, R_\text{MH}) \gets \Call{Rand}{M}$
    \For{$i \gets (j + 1):K$} \Comment{Loop over chains.}
     \If {$a[i] > 0$} $Q[i] \gets Q[a[i]]$ \textbf{else} \Comment{Copy, else run MCMC.}
      \State $Q^-[i] \gets$ \Call{Mcmc}{$Q[i], M, R_\text{MCMC}$}
      \If{$i = j + 1$} \label{line:maxcouplelower1}
       \State $m \gets 1$ \Comment{Maximally couple row~$j + 1$ with row~1.}
      \Else \Comment{$i > j + 1$: Couple with some row $\in (j + 1):(i - 1)$.}
       \State $m \gets \Call{MinInd}{Q^-, j + 1, i}$ \Comment{Choose closest row.}
      \EndIf
      \State $Q^*[i] \gets \Call{MaxCouple}{Q^-[m], Q^*[m], Q^-[i], r}$ \label{line:maxcouplelower0}
      \State $Q[i] \gets \Call{MhTest}{Q^-[i], Q^*[i], R_\text{MH}}$
      \If {$Q[i] = Q[m]$} \Comment{$m \gets$ earliest coalescence row for $i$.}
       \State \textbf{if} $a[m] > 1$ \textbf{then} $m \gets a[m]$ \textbf{end if}
       \State $a[i] \gets m$ \Comment{Record coalescence of row~$i$ with earliest.}
       \State \textbf{if} $m > 1$ \textbf{then} $a[a = i] \gets m$ \textbf{end if}
      \EndIf \label{line:maxcouplelower2}
     \EndIf
    \EndFor \Comment{$i$ loop (rows of Fig.~\ref{fig:sampleset})}
   \EndFor \Comment{$j_1$ loop over sub-blocks}
  \EndFor \Comment{$j$ loop (columns of Fig.~\ref{fig:sampleset})}
  \State Error $\gets$ Error \textbf{or} $a[K] \ne 1$; $Q[1] \gets Q_1[K, J]$\Comment{Check $K$; restore~1.}
 \end{algorithmic}
\end{algorithm}

We insert one Metropolis--Hastings update, with maximal coupling and a test for coalescence, after every $M$ MCMC iterations in each cell of Fig.~\ref{fig:sampleset}, where $M$ is a divisor of the block length~$B$ (see lines \ref{line:maxcoupleupper1}--\ref{line:maxcoupleupper2} in Algorithm~\ref{alg:replaceupper} and \ref{line:maxcouplelower1}--\ref{line:maxcouplelower2} in Algorithm~\ref{alg:replacelower}).  In row~1 of Fig.~\ref{fig:sampleset}, jump vectors are simulated freely from a solid spherical uniform distribution (Algorithm~\ref{alg:replaceupper}, line~\ref{line:freejump}).  In each subsequent row~$Y$, each jump vector is calculated from that of the row~$X$ with which it is maximally coupled, according to function \textsc{MaxCouple} of Algorithm~\ref{alg:maximal}, and there is no extra random input (Algorithm~\ref{alg:replaceupper}, line~\ref{line:maxcoupleupper0}, and Algorithm~\ref{alg:replacelower}, line~\ref{line:maxcouplelower0}).

Three versions of the result column-vector $Q$ are recorded through the algorithm: the pre-jump~$Q^-$, the post-jump~$Q^*$, and the final version~$Q$ which is equal to either $Q^*$ or~$Q^-$ depending on whether the jump is accepted or rejected in the M--H test.  These are overwritten in each random-number sub-block of $M$ MCMC iterations (sub-column of Fig.~\ref{fig:sampleset}).  After each sub-block of row~1 of Fig.~\ref{fig:sampleset}, all three versions are stored for later use in row-vectors $Q_1^-$, $Q_1^*$ and~$Q_1$, to save having to recompute them in the lower triangle of Fig.~\ref{fig:sampleset}.

\section{Examples} \label{sec:examplesimple}

\subsection{Two-state discrete process} \label{sec:exampletwostate}

\begin{figure}[ht]
 \centering
 %\includegraphics[width=0.9\textwidth, trim=1.5cm 9.3cm 2cm 2.3cm,
 % clip=true]{figtransitionsimple}
 \includegraphics[width=0.2\textwidth, trim=2.8cm 9.3cm 8cm 2.3cm,
  clip=true]{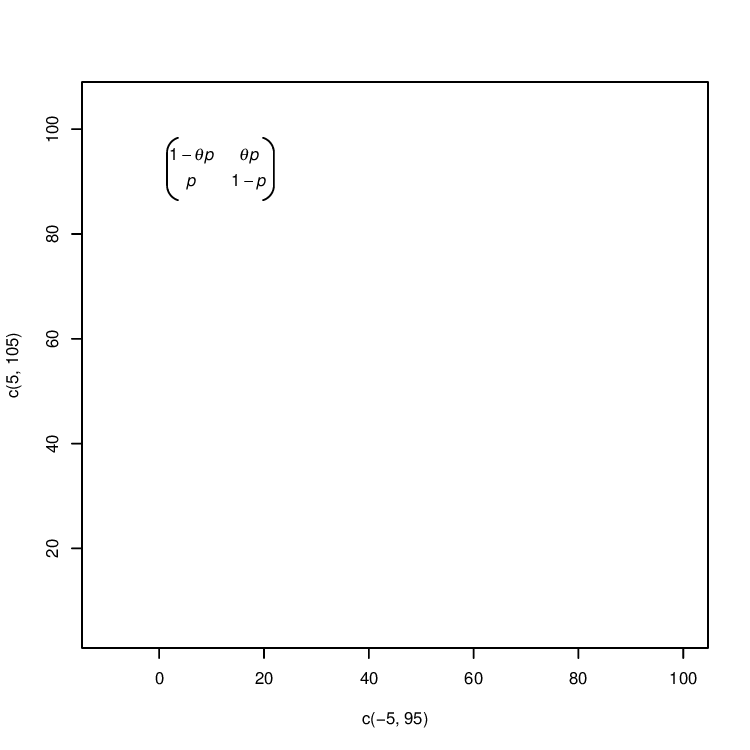}
 \caption[Markov transition matrix]
 {Markov transition matrix for the two-state process.  Parameter values are $\theta = 1 / 9$ and $p = 0.1$} \label{fig:transitionsimple}
\end{figure}

We illustrate the methodology of section~\ref{sec:unbiasedperfect} on a two-state process with the transition matrix shown in Fig.~\ref{fig:transitionsimple} and parameters $\theta$ and~$p$ satisfying $0 \le \theta \le 1$ and $0 < p \le \half$.

Starting points for different chains are generated independently, with starts from states 1 and~2 having probabilities (0.5, 0.5).  We set $\theta = 1 / 9$ and $p = 0.1$.  Chains converge to  the stationary distribution ($1 / (1 + \theta)$, $\theta / (1 + \theta)$) = (0.9, 0.1).  The probability $p$ regulates the speed of convergence.  The determinant of the transition matrix, which we use below, is $\Delta = 1 - p - \theta p$ = $8 / 9$.

Chains $X$ and~$Y$ are coupled by generating a single sequence $S_1$, $S_2$, \ldots\ of independent uniform (0, 1) random numbers, and using $S_i$ to transition from $X_{i - 1}$ to~$X_i$ and from $Y_{i - 2}$ to~$Y_{i - 1}$.  At time step~$i$, if chain $X$ is in state~1, it moves to state~2 if $S_i > 1 - \theta p$.  If it is in state~2, it moves to state~1 if $S_i < p$.  Chain~$Y$ uses $S_{i + 1}$ instead of~$S_i$, and does not use~$S_1$.

In this example, the probability that $X_i \ne Y_{i - 1}$ for $i \ge 1$ can be found analytically.  For $i = 1$ it is equal to $\half$ because $Y_0$ is sampled independently of $X_1$ and has a probability of $\half$ of being equal to $X_1$.  For $i > 1$, if $X_{i - 1} \ne Y_{i - 2}$, $X_i$ and $Y_{i - 1}$ will coalesce to state~1 if $S_i < p$, and to state~2 if $S_i > 1 - \theta p$.  The probability of not coalescing at step~$i$ is the determinant $\Delta$ calculated above, and the overall probability of failing to coalesce by step~$i$ is $\half \Delta^{i - 1}$.  This probability takes the values 0.31 at $i = 5$, 0.053 at $i = 20$, $4.3 \times 10^{-6}$ at $i = 100$, and $1.5 \times 10^{-26}$ at $i = 500$.

\begin{figure}[ht]
 \centering
 \includegraphics[width=0.8\textwidth, trim=0cm 0.5cm 1cm 2cm,
  clip=true]{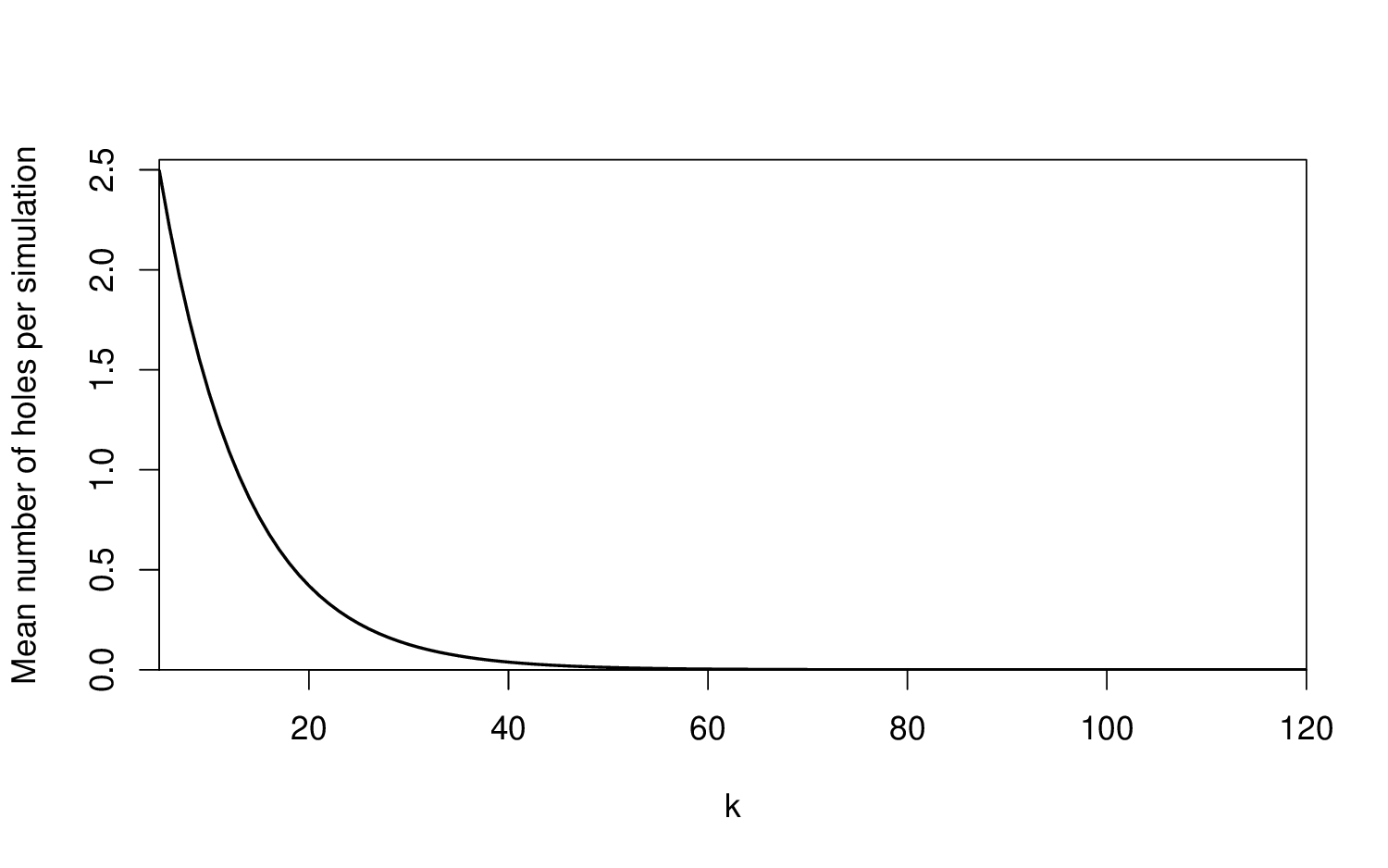}
 \caption[Proportion of holes in the two-state process]
 {Sample mean number of holes per simulation of the two-state process, as a function of burn-in time~$k$ for $k \ge 5$.  No holes were observed for $k \ge 106$} \label{fig:numberofholes}
\end{figure}

We ran one million independent simulations of $X$ and~$Y$, and applied the methodology of section~\ref{sec:unbiasedperfect}, varying the number of burn-in iterations $k$ from 5 to~120.  For $k \ge 106$, $X$ and~$Y$ coalesced by iteration $k + 1$ in all simulations, so that no holes remained.  Table~\ref{tab:results2state} shows the results for selected values of~$k$.  The proportion of holes in the results, as a function of~$k$, is plotted in Fig.~\ref{fig:numberofholes}.  At $k = 50$, the number of holes becomes low, although the sample standard deviation of the sum of the weights (which take values $\pm 1$) of the sample points is still substantially greater than the converged value of~0.3.  The contrast of columns 2 and~3 in Table~\ref{tab:results2state} shows the effectiveness of the unbiased sampling methodology, which produces an estimated stationary probability of being in state~1 that does not differ significantly from the true stationary value of~0.9.

\begin{table}[th]
 \caption[Results for the two-state process] {Results from samples of 1 million independent simulations of the two-state process, for various lengths of the burn-in time~$k$.  Other columns are the unadjusted proportion of values of $X_k$ in state~1; the adjusted proportion using values $X_k$, \ldots, $X_\tau$ and $Y_k$, \ldots, $Y_{\tau - 1}$ and weights of $\pm 1$ from the theory in section~\ref{sec:unbiasedperfect}; the proportion of strings of length $\nu > 1$ in the sample; the total number of holes as a proportion of the sample size; and the sample standard deviation of the sum of weights $w_i = \pm 1$ of string elements (from all strings, whether $\nu = 1$ or $\nu > 1$) that are in state~1}  \label{tab:results2state}
 \centering
 \begin{tabular}{l r r l l r}
 \hline
 \bigstrut[t] \phantom{00}$k$ & Unadj. & Adj. & $\nu > 1$ & Holes & S.d. \\
 \hline
 \bigstrut[t] \phantom{00}5 & 0.677422 & 0.899944 & 0.2776 & 2.4944 & 6.6841 \\
 \phantom{0}10 & 0.776260 & 0.902078 & 0.1541 & 1.3817 & 4.9703 \\
 \phantom{0}20 & 0.861767 & 0.900215 & 0.0475 & 0.4204 & 2.7460 \\
 \phantom{0}50 & 0.898480 & 0.899104 & 0.0013 & 0.0118 & 0.5454 \\
 100 & 0.899329 & 0.899335 & $4 \times 10^{-6}$ & $1.2 \times 10^{-5}$ &
  0.3010 \\
           110 & 0.899829 & 0.899829 & 0      & 0      & 0.3002 \\
 \hline
 \end{tabular}
\end{table}
 
When perfect simulation is desired, $k$ should be set before the results are generated, to a value high enough to make the probability of any lack of coalescence after $k$ iterations very small.  It would break the rules to allow the final simulations to influence the choice of~$k$.

If the methods of section~\ref{sec:algstructure} were applied to this example, we might decide on a sample-set size of $K = 20$ and a block length of $B = 25$.  This would yield an effective value of $k$ of about $KB = 500$, and probabilities of lack of coalescence after $K + 1$ blocks approximately equal to $10^{-26}$ as calculated above for an individual simulation, and $10^{-20}$ in a set of $10^6$ simulations.  It is straightforward to show that raising the transition matrix to the power $B$ is equivalent to replacing $p$ by $p (1 - \Delta^B) / (1 - \Delta)$, and then that if $X_{K B}$ and $X_{(K + 1) B}$ both follow the stationary distribution, their correlation coefficient is $\Delta^B$ = 0.0526, thereby quantifying analytically the serial correlation of perfect sample points in the same sample set in this example.

\subsection{Normal distribution} \label{sec:examplenormal}

We applied the methodology of sections \ref{sec:algstructure} and~\ref{sec:maximal} to the simulation of a standard normal distribution in $d$ dimensions, using random-walk MCMC with an internal Metropolis--Hastings algorithm.  We used a $N(0, \sigma I)$ distribution for the jumps in the internal MCMC, with $\sigma = 2 / \sqrt d\,$, where $I$ denotes the $d$-dimensional identity matrix.  For the maximal coupling steps, we set the radius of the solid spherical uniform jump distribution to $r = 3$.

The $K$ chains in a sample set were coupled by using the same relative jumps from each chain's current position.  Thereby, two chains stayed the same distance apart unless a jump was accepted for one chain but not the other.  Coalescence relied on different chains ``piling up'' around the effective boundary of the sample space, beyond which, even though the space was of infinite extent, the M--H probability of acceptance was low.  M--H rejection could stop one chain from going through the boundary, allowing another chain that had been further from that boundary to catch up with it.  The internal M--H acceptance test used the same uniform (0, 1) random variable for all chains.

Due to the relative clumsiness with which coalescence is achieved in random-walk MCMC, maximal coupling, with its own, separate M--H update, was conducted after every MCMC step, i.e., $M = 1$ in section~\ref{sec:maximal}.  Hamiltonian Monte Carlo, although beyond the scope of this paper, performs this and similar tasks more efficiently than random-walk MCMC (see section~\ref{sec:discussion} below).

All chains began by sampling each of the $d$ coordinates independently from a uniform $(-6, 6)$ distribution.  As suggested in section~\ref{sec:algstructure},  we set the block length $B$ from an exploratory analysis, to aim for a probability of 0.1 that two chains would fail to coalesce within $B$ iterations.  We set the sample-set size $K$ to 20, with the aim of achieving a probability that two chains would fail to coalesce after $K$ blocks (or $K B$ MCMC iterations) of roughly $10^{-20}$.

Results are shown in Table~\ref{tab:resultsnormal}.  The maximum number of blocks for a chain to coalesce to the chain below it in Fig.~\ref{fig:sampleset} for any of the runs is 9, and does not approach the sample-set size $K = 20$.  Therefore, holes do not occur in any of these results, and the algorithms have generated perfect samples without any need to process holes.

\begin{table}[th]
 \caption[Results for the standard normal distribution] {Results for simulation of a standard normal distribution by random-walk MCMC, using the methodology of sections~\ref{sec:algstructure} and~\ref{sec:maximal}, for various numbers of dimensions~$d$.  Other columns are the chosen block length $B$, the total number of chains or sample points $N$, the sample mean and maximum number of blocks for each chain to coalesce with the one below it in Fig.~\ref{fig:sampleset}, and the sample coefficient $\rho$ of the serial correlation which occurs by design between points in the same sample set.  The sample-set size $K$ was set to 20 in all simulations, so that the effective number of iterations run for each chain was $20 B$ and the number of independent sample sets was $N / 20$}  \label{tab:resultsnormal}
 \centering
 \begin{tabular}{r r l l c r}
 \hline
 \bigstrut[t] $d$ & $B$ & $N$ & Mean & Max. & $\rho \phantom{.00000}$ \\
 \hline
 \bigstrut[t] 1 & 5 & $10^7$ & 1.111 & 9 & $0.00940$ \\
 2 & 10 & $10^6$ & 1.085 & 6 & $0.00263$ \\
 5 & 25 & $10^6$ & 1.107 & 7 & $0.00201$ \\
 10 & 95 & $10^6$ & 1.105 & 7 & $-0.00055$ \\
 15 & 425 & $10^6$ & 1.114 & 9 & $0.00055$ \\
 20 & 3500 & $10^5$ & 1.141 & 9 & $-0.00084$ \\
 \hline
 \end{tabular}
\end{table}

Geometric decay in the number of blocks to coalesce is followed tolerably closely, as indicated by the sample means and maxima.  The means remain close to the target of about $1 / 0.9 \approx 1.11$; i.e., 90\% of non-coalesced chains coalesce within one further block.  Detailed inspection of the numbers for $d = 15$ and~20, however, indicated some departure in this desideratum.  For $d = 15$, the numbers of chains requiring 3, 4, 5, 6, 7, 8 and 9 blocks to coalesce were 11341, 2088, 424, 98, 17, 8 and 1, which imply that, beyond 3 blocks, only 80\% or fewer were coalescing in one further block.  For $d = 20$, the numbers were 1532, 316, 56, 6, 3, 1 and 1, which imply much the same.  The required block length~$B$ to achieve these rates of coalescence also increased with~$d$ at a rate greater than exponential.  To us these problems indicate that $d = 20$ is close to the limit at which random-walk MCMC can function effectively.  As a comparison, for $d = 5$ the numbers of chains requiring 2, 3, 4, 5, 6 and 7 blocks were 95896, 5082, 397, 36, 1 and 1, showing that, in all but the final category, more than 90\% of non-coalesced chains coalesced in one further block.

Serial correlation of chains in the same sample set (final column of Table~\ref{tab:resultsnormal}) is a design feature of the algorithm in section~\ref{sec:algstructure}, introduced to enable more perfect samples to be generated for very little extra computation.  The sample correlation was statistically significant (about 30 standard deviations) at $d = 1$, for which the block length $B$ was only~5.  It was marginally significant (about 4~s.d.) at $d = 2$ and $d = 5$.  It was not statistically significant for $d \ge 10$.

\section{Discussion} \label{sec:discussion}

We have demonstrated how unbiased simulation can be converted into perfect simulation with an arbitrarily small probability that complications will arise, which we have set to roughly $10^{-20}$.  To handle cases where complications nevertheless do arise, we have developed the theory of holes in section~\ref{sec:relaxed}, and shown the effect that holes have on the standard deviation of a parameter estimator in one simple example (Table~\ref{tab:results2state}).  In our other example (section~\ref{sec:examplenormal} and Table~\ref{tab:resultsnormal}), the number of blocks for neighbouring chains to coalesce and enable perfect simulation was less than the limiting number of blocks $K = 20$ by a comfortable margin in all of our runs.  Hence, holes did not arise.

The nonzero probability of occurrence of holes in a sample appeases the theoretical limitation that, unless a property such as monotonicity is available, a general, strictly perfect simulation scheme must visit every possible state of a process (see section~\ref{sec:intro}).  In that sense, the possibility of holes is to be welcomed rather than cursed.  In practice, due to their low theoretical probability, the occurrence of holes in a sample would suggest that some feature of the likelihood function has been overlooked during theoretical examination.  This in turn may highlight the need to redesign the internal MCMC algorithm that is called by our algorithms~\ref{alg:practical}, \ref{alg:replaceupper} and~\ref{alg:replacelower}.

Our algorithms in section~\ref{sec:algorithm} introduce a trade-off between computational effort and serial correlation within a sample set.  Despite this trade-off, all of the generated points remain perfect sample points, and different sample sets, each containing about 20~points, remain completely independent.  We view serial correlation of perfect samples as a smaller price to pay than the disagreeable behaviour that can arise in unbiased sampling with low to moderate burn-in times $k$ (see sections~\ref{sec:intro}, \ref{sec:recap} and~\ref{sec:exampletwostate}).  The computation per final sample point in our algorithm is about the same as that of unbiased simulation with $k$ set to our block length $B$.  Unbiased simulation runs the same block of random numbers on two chains, to generate one point, with relatively little extra computation when chains don't coalesce after one block.  Our algorithm runs a minimum of $2 K$ blocks of random numbers (marked red in Fig.~\ref{fig:sampleset}) to generate $K$ points, again with a little extra when coalescence takes more than one block.

It can be argued that perfect simulation is beyond price, given the very high uncertainty around MCMC convergence that exists with current tests (see section~\ref{sec:intro}).  We note that any failure of our methodology to produce perfect sample points is immediately clear from the associated lack of coalescence after $K$ blocks.  Such clarity is not a feature of current MCMC convergence tests.

We conclude with a discussion of Hamiltonian Monte Carlo (HMC), a technique whose combination with the developments in this paper offers exceptional performance and almost complete achievement of the objective of replacing uncertain MCMC convergence tests by perfect simulation, for continuous processes with differentiable likelihoods.  HMC was developed by~\citet{duane_hybrid_1987} and is discussed in detail by~\citet{neal_probabilistic_1993, neal_bayesian_1995, brooks_mcmc_2011}.  Originally known as Hybrid Monte Carlo, HMC alternates random and deterministic phases, thereby exploring the sample space much more effectively than older MCMC methods that are based on random walks \citep[see][sec. 5.3.3]{brooks_mcmc_2011}.

Originally, each instance of the deterministic phase, termed a ``trajectory'', ran for a fixed length of notional time.  Advanced HMC algorithms, of which the first was the No-U-Turn Sampler (NUTS) \citep{hoffman_no-u-turn_2014}, adjust the trajectory length according to the curvature of the likelihood function.  \citet{leigh_design_2022} developed one completely new advanced HMC algorithm, and another algorithm that is an improved version of NUTS.

Unbiased sampling was applied to original HMC by~\citet{heng_unbiased_2019}, using the observation that coupled trajectories with different starting points are drawn closer together when their trajectory time-span is equal to one quarter of the period of a complete orbit.  \citet{leigh_design_2022} applied unbiased sampling to the advanced HMC algorithms by coupling the selection of a trajectory's destination point, as a random proportion of the traversal time from the back endpoint to the forward one.  They achieved perfect sampling from a wide range of distributions, including long-tailed $t$-distributions and Gaussian mixtures, up to 100 dimensions.  \citet{bou-rabee_coupling_2020} applied coupling to HMC and included varying the initial velocities of coupled trajectories, whereas other couplings set the initial velocities equal.

Unlike random-walk MCMC, we know of no limit to the number of dimensions in problems to which HMC can be applied.

\section*{Declarations}

No funding was received for conducting this study.

The authors have no relevant financial or non-financial interests to disclose.

The authors have no conflicts of interest to declare that are relevant to the content of this article.

All authors certify that they have no affiliations with or involvement in any organization or entity with any financial interest or non-financial interest in the subject matter or materials discussed in this manuscript.

The authors have no financial or proprietary interests in any material discussed in this article.

Code for the examples in section~\ref{sec:examplesimple} is available at \url{https://github.com/George-Leigh/PerfectSimulation}, written in R \citep{r_core_team_r_2023}.

\section*{Authors' note}

This paper is a restructured and expanded version of some of the material
 contained in \citet{leigh_design_2022}.  This paper contains new theory for
 maximal coupling, which we believe to be superior to the congruential
 coupling (called a ``rounding step'') of the original paper.  The example
 applications of the methodology are also new.

\bibliographystyle{chicago}
\bibliography{GeorgeLibrary}% Common bib file if required, the content of .bbl file can be included here once bbl is generated

\begin{thebibliography}{}

\bibitem[\protect\citeauthoryear{Aldous and Diaconis}{Aldous and
  Diaconis}{1987}]{aldous_strong_1987}
Aldous, D. and P.~Diaconis (1987).
\newblock Strong uniform times and finite random walks.
\newblock {\em Adv. Appl. Math.\/}~{\em 8\/}(1), 69--97.

\bibitem[\protect\citeauthoryear{Bellman}{Bellman}{1955}]{bellman_review_1955}
Bellman, R. (1955).
\newblock Review of {Transactions} of the {Symposium} on {Computing},
  {Mechanics}, {Statistics}, and {Partial} {Differential} {Equations}, {Vol.}
  {II}.
\newblock {\em J. Amer. Stat. Assoc.\/}~{\em 50\/}(272), 1357--1359.

\bibitem[\protect\citeauthoryear{Bou-Rabee, Eberle, and Zimmer}{Bou-Rabee
  et~al.}{2020}]{bou-rabee_coupling_2020}
Bou-Rabee, N., A.~Eberle, and R.~Zimmer (2020).
\newblock Coupling and convergence for {Hamiltonian} {Monte} {Carlo}.
\newblock {\em Ann. Appl. Probab.\/}~{\em 30\/}(3), 1209--1250.
\newblock Publisher: Institute of Mathematical Statistics.

\bibitem[\protect\citeauthoryear{Brooks, Gelman, Jones, and Meng}{Brooks
  et~al.}{2011}]{brooks_handbook_2011}
Brooks, S., A.~Gelman, G.~Jones, and X.-L. Meng (2011).
\newblock {\em Handbook of {Markov} {Chain} {Monte} {Carlo}}.
\newblock Chapman and Hall.

\bibitem[\protect\citeauthoryear{Craiu and Meng}{Craiu and
  Meng}{2011}]{craiu_perfection_2011}
Craiu, R.~V. and X.-L. Meng (2011).
\newblock Perfection within reach: {Exact} {MCMC} sampling.
\newblock In S.~Brooks, A.~Gelman, G.~Jones, and X.-L. Meng (Eds.), {\em
  Handbook of {Markov} {Chain} {Monte} {Carlo}}, pp.\  199--226. Chapman and
  Hall.

\bibitem[\protect\citeauthoryear{Duane, Kennedy, Pendleton, and Roweth}{Duane
  et~al.}{1987}]{duane_hybrid_1987}
Duane, S., A.~D. Kennedy, B.~J. Pendleton, and D.~Roweth (1987).
\newblock Hybrid {Monte} {Carlo}.
\newblock {\em Phys. Lett. B\/}~{\em 195\/}(2), 216--222.

\bibitem[\protect\citeauthoryear{Fill}{Fill}{1998}]{fill_interruptible_1998}
Fill, J.~A. (1998).
\newblock An interruptible algorithm for perfect sampling via {Markov} chains.
\newblock {\em Ann. Appl. Probab.\/}~{\em 8\/}(1), 131--162.

\bibitem[\protect\citeauthoryear{Gelman and Shirley}{Gelman and
  Shirley}{2011}]{gelman_inference_2011}
Gelman, A. and K.~Shirley (2011).
\newblock Inference and monitoring convergence.
\newblock In S.~Brooks, A.~Gelman, G.~Jones, and X.-L. Meng (Eds.), {\em
  Handbook of {Markov} {Chain} {Monte} {Carlo}}, pp.\  163--174. Chapman and
  Hall.

\bibitem[\protect\citeauthoryear{Glynn and Rhee}{Glynn and
  Rhee}{2014}]{glynn_exact_2014}
Glynn, P.~W. and C.-H. Rhee (2014).
\newblock Exact estimation for {Markov} chain equilibrium expectations.
\newblock {\em J. Appl. Probab.\/}~{\em 51\/}(A), 377--389.

\bibitem[\protect\citeauthoryear{Griffeath}{Griffeath}{1975}]{griffeath_maximal_1975}
Griffeath, D. (1975).
\newblock A maximal coupling for {Markov} chains.
\newblock {\em Z. Wahrscheinlichkeitstheorie Verwandte Geb.\/}~{\em 31\/}(2),
  95--106.

\bibitem[\protect\citeauthoryear{Hastings}{Hastings}{1970}]{hastings_monte_1970}
Hastings, W.~K. (1970).
\newblock Monte {Carlo} sampling methods using {Markov} chains and their
  applications.
\newblock {\em Biometrika\/}~{\em 57\/}(1), 97--109.

\bibitem[\protect\citeauthoryear{Heng and Jacob}{Heng and
  Jacob}{2019}]{heng_unbiased_2019}
Heng, J. and P.~E. Jacob (2019).
\newblock Unbiased {Hamiltonian} {Monte} {Carlo} with couplings.
\newblock {\em Biometrika\/}~{\em 106\/}(2), 287--302.

\bibitem[\protect\citeauthoryear{Hoffman and Gelman}{Hoffman and
  Gelman}{2014}]{hoffman_no-u-turn_2014}
Hoffman, M.~D. and A.~Gelman (2014).
\newblock The {No}-{U}-{Turn} {Sampler}: {Adaptively} setting path lengths in
  {Hamiltonian} {Monte} {Carlo}.
\newblock {\em J. Mach. Learn. Res.\/}~{\em 15}, 1593--1623.

\bibitem[\protect\citeauthoryear{Huber}{Huber}{2016}]{huber_perfect_2016}
Huber, M.~L. (2016).
\newblock {\em Perfect {Simulation}}.
\newblock Boca Raton, FL: CRC Press.

\bibitem[\protect\citeauthoryear{Jacob, O’Leary, and Atchadé}{Jacob
  et~al.}{2020}]{jacob_unbiased_2020}
Jacob, P.~E., J.~O’Leary, and Y.~F. Atchadé (2020).
\newblock Unbiased {Markov} chain {Monte} {Carlo} methods with couplings.
\newblock {\em J. R. Stat. Soc. Ser. B. Stat. Methodol.\/}~{\em 82\/}(3),
  543--600.

\bibitem[\protect\citeauthoryear{Johnson}{Johnson}{1998}]{johnson_coupling-regeneration_1998}
Johnson, V.~E. (1998).
\newblock A coupling-regeneration scheme for diagnosing convergence in {Markov}
  chain {Monte} {Carlo} algorithms.
\newblock {\em J. Am. Stat. Assoc.\/}~{\em 93\/}(441), 238--248.
\newblock Publisher: Taylor \& Francis \_eprint:
  https://www.tandfonline.com/doi/pdf/10.1080/01621459.1998.10474105.

\bibitem[\protect\citeauthoryear{Leigh and Northrop}{Leigh and
  Northrop}{2022}]{leigh_design_2022}
Leigh, G.~M. and A.~R. Northrop (2022).
\newblock Design of {Hamiltonian} {Monte} {Carlo} for perfect simulation of
  general continuous distributions.
\newblock Technical Report arXiv:2212.12140, arXiv.

\bibitem[\protect\citeauthoryear{Liu}{Liu}{2004}]{liu_monte_2004}
Liu, J.~S. (2004).
\newblock {\em Monte {Carlo} {Strategies} in {Scientific} {Computing}}.
\newblock Springer {Series} in {Statistics}. New York, NY: Springer.

\bibitem[\protect\citeauthoryear{Lovász and Winkler}{Lovász and
  Winkler}{1995}]{lovasz_exact_1995}
Lovász, L. and P.~Winkler (1995).
\newblock Exact mixing in an unknown {Markov} chain.
\newblock {\em Electron. J. Comb.\/}~{\em 2}, article R15.

\bibitem[\protect\citeauthoryear{Metropolis, Rosenbluth, Rosenbluth, Teller,
  and Teller}{Metropolis et~al.}{1953}]{metropolis_equations_1953}
Metropolis, N., A.~W. Rosenbluth, M.~N. Rosenbluth, A.~H. Teller, and E.~Teller
  (1953).
\newblock Equations of state calculations by fast computing machines.
\newblock {\em J. Chem. Phys.\/}~{\em 21}, 1087--1092.

\bibitem[\protect\citeauthoryear{Mira, Møller, and Roberts}{Mira
  et~al.}{2001}]{mira_perfect_2001}
Mira, A., J.~Møller, and G.~O. Roberts (2001).
\newblock Perfect slice samplers.
\newblock {\em J. R. Stat. Soc. Ser. B. Stat. Methodol.\/}~{\em 63\/}(3),
  593--606.

\bibitem[\protect\citeauthoryear{Murdoch and Green}{Murdoch and
  Green}{1998}]{murdoch_exact_1998}
Murdoch, D.~J. and P.~J. Green (1998).
\newblock Exact sampling from a continuous state space.
\newblock {\em Scand. J. Stat.\/}~{\em 25\/}(3), 483--502.

\bibitem[\protect\citeauthoryear{Neal}{Neal}{1993}]{neal_probabilistic_1993}
Neal, R.~M. (1993).
\newblock Probabilistic {Inference} {Using} {Markov} {Chain} {Monte} {Carlo}
  {Methods}.
\newblock Research report, Department of Computer Science, University of
  Toronto.

\bibitem[\protect\citeauthoryear{Neal}{Neal}{1995}]{neal_bayesian_1995}
Neal, R.~M. (1995).
\newblock {\em Bayesian {Learning} for {Neural} {Networks}}.
\newblock {PhD} thesis, University of Toronto, Graduate Department of Computer
  Science.

\bibitem[\protect\citeauthoryear{Neal}{Neal}{2011}]{brooks_mcmc_2011}
Neal, R.~M. (2011).
\newblock {MCMC} using {Hamiltonian} dynamics.
\newblock In S.~Brooks, A.~Gelman, G.~Jones, and X.-L. Meng (Eds.), {\em
  Handbook of {Markov} {Chain} {Monte} {Carlo}}, pp.\  113--162. Chapman and
  Hall.

\bibitem[\protect\citeauthoryear{Propp and Wilson}{Propp and
  Wilson}{1996}]{propp_exact_1996}
Propp, J.~G. and D.~B. Wilson (1996).
\newblock Exact sampling with coupled {Markov} chains and applications to
  statistical mechanics.
\newblock {\em Random Struct. Algorithms\/}~{\em 9\/}(1–2), 223--252.

\bibitem[\protect\citeauthoryear{{R Core Team}}{{R Core
  Team}}{2023}]{r_core_team_r_2023}
{R Core Team} (2023).
\newblock {\em R: {A} language and environment for statistical computing}.
\newblock Vienna: R Foundation for Statistical Computing.

\bibitem[\protect\citeauthoryear{Roy}{Roy}{2020}]{roy_convergence_2020}
Roy, V. (2020, March).
\newblock Convergence diagnostics for {Markov} chain {Monte} {Carlo}.
\newblock {\em Annu. Rev. Stat. Appl.\/}~{\em 7}, 387--412.

\bibitem[\protect\citeauthoryear{van~den Boom, Jasra, De~Iorio, Beskos, and
  Eriksson}{van~den Boom et~al.}{2022}]{van_den_boom_unbiased_2022}
van~den Boom, W., A.~Jasra, M.~De~Iorio, A.~Beskos, and J.~G. Eriksson (2022).
\newblock Unbiased approximation of posteriors via coupled particle {Markov}
  chain {Monte} {Carlo}.
\newblock {\em Stat. Comput.\/}~{\em 32\/}(3), 36.

\bibitem[\protect\citeauthoryear{Wilson}{Wilson}{2000}]{wilson_how_2000}
Wilson, D.~B. (2000).
\newblock How to couple from the past using a read-once source of randomness.
\newblock {\em Random Struct. Algorithms\/}~{\em 16\/}(1), 85--113.

\end{thebibliography}
%\input sn-article.bbl

% Default
%\input sn-sample-bib.tex%

\end{document}